\documentclass[aps,prc,reprint,amsmath,amssymb,showpacs]{revtex4-1}
\usepackage{CJK}
\usepackage{graphicx}
\usepackage{dcolumn}
\usepackage{bm}

\begin{document}

\title{Yrast band of $^{109}$Ag described by tilted axis cranking covariant density functional theory with a separable pairing force}
\author{Y. K. Wang}
\affiliation{State Key Laboratory of Nuclear Physics and Technology, School of Physics, Peking University, Beijing 100871, China}

\date{\today}
\begin{abstract}
  A separable form of the Gogny pairing force is implemented in tilted axis cranking covariant density functional theory for the description of rotational bands in open shell nuclei.
  The developed method is used to investigate the yrast sequence of $^{109}$Ag for an example.
  The experimental energy spectrum, angular momenta, and electromagnetic transition probabilities are well reproduced by taking into account pairing correlations with the separable pairing force.
  An abrupt transition of the rotational axis from the long-intermediate plane to the long-short one is obtained and discussed in detail.
\end{abstract}

\maketitle

\date{today}

\section{Introduction}
The most common rotational bands in nuclei are built on states with a substantial quadrupole deformation and axial symmetry.
They show strong electric quadrupole ($E2$) transitions between the rotational states.
Such bands are usually well interpreted as a coherent collective rotation of many nucleons around an axis perpendicular to the symmetry axis of the deformed density distribution.
However, since the discovery of many rotational-like sequences in weakly deformed nuclei in 1990s~\cite{Hubel2005Prog.Part.Nucl.Phys.1}, the rotation of nuclei around a tilted axis has attracted a lot of attention in both theoretical and experimental studies~\cite{Frauendorf2001Rev.Mod.Phys.463}.
In this case the rotational axis is tilted with respect to the principal axes of the density distribution in order to account for the fact that the nucleus is composed of a deformed core and nucleons carrying a quantized amount of angular momentum.
Several typical examples of the tilted axis rotation in nuclei include the high-$K$ bands that give rise to $K$ isomerism~\cite{Walker1999Nature35} in axially deformed nuclei, magnetic rotation in weakly deformed nuclei~\cite{Frauendorf2001Rev.Mod.Phys.463,Clark2000Annu.Rev.Nucl.Part.Sci.1,Meng2013Front.Phys.55}, chiral rotation in triaxial nuclei~\cite{Frauendorf1997Nucl.Phys.A131,Frauendorf2001Rev.Mod.Phys.463,Meng2006Phys.Rev.C037303,Meng2010JPhysG.37.64025}, etc.

The tilted axis rotation was first proposed within the tilted axis cranking (TAC) model based on a Nilsson mean field~\cite{Frauendorf1993Nucl.Phys.A259}.
Later on, the quality of the TAC approximation was examined in comparison with the quantum particle rotor model (PRM)~\cite{Frauendorf1996Z.Phys.A263}.
Since then, the TAC model has become a powerful tool to describe nuclear tilted axis rotations.
Moreover, in this model it is relatively easy to construct the vector diagrams of angular momentum composition, because this approach is based on a classical picture of rotation.
However, due to the numerical complexity of the TAC model, most of the applications are based on
the pairing-plus-quadrupole model and the Strutinsky shell-correction model~\cite{Frauendorf2001Rev.Mod.Phys.463,Frauendorf2000Nucl.Phys.A115}.

In recent years, the TAC approaches based on relativistic~\cite{Madokoro2000Phys.Rev.C61301,Peng2008Phys.Rev.C24313,Zhao2011Phys.Lett.B181} and non-relativistic~\cite{Olbratowski2002ActaPhys.Pol.B389,Olbratowski2004Phys.Rev.Lett.52501,Olbratowski2006Phys.Rev.C54308} density functional theories have been developed and achieved great successes~\cite{Meng2013Front.Phys.55,Meng2016Phys.Scr53008}, similarly as the principal axis cranking density functional theory~\cite{Afanasjev2000Nucl.Phys.A196-244,Afanasjev2013Phys.Rev.C014320}.
These self-consistent methods are based on more realistic two-body interactions and, thus, can be used to study the nuclear rotational excitations on a more fundamental level by including all important effects, such as core polarization and nuclear currents~\cite{Meng2013Front.Phys.55,Olbratowski2004Phys.Rev.Lett.52501,Zhao2011Phys.Rev.Lett.122501}.
In particular, the tilted axis cranking covariant density functional theory (TAC-CDFT) provides a consistent description of currents and time-odd fields, and the included nuclear magnetism~\cite{Koepf1989Nucl.Phys.A61} plays an important role in the description of nuclear rotations~\cite{Konig1993Phys.Rev.Lett.3079,Afanasjev2000Phys.Rev.C31302,Afanasjev2010Phys.Rev.C34329,Liu2012Sci.ChinaPhys.Mech.Astron.2420}.
So far, the two-dimensional TAC-CDFT has been successfully used to describe the magnetic rotational bands~\cite{Zhao2011Phys.Lett.B181, Yu2012Phys.Rev.C024318, Peng2015Phys.Rev.C044329}, antimagnetic rotational bands~\cite{Zhao2011Phys.Rev.Lett.122501, Zhao2012Phys.Rev.C054310, Peng2015Phys.Rev.C044329}, linear alpha cluster bands~\cite{Zhao2015Phys.Rev.Lett022501}, etc., and has demonstrated high predictive power~\cite{Meng2013Front.Phys.55,Meng2016Phys.Scr53008}.

Very recently, pairing correlations have been considered self-consistently in the TAC-CDFT by solving the corresponding relativistic Hartree Bogoliubov (RHB) equations with a monopole pairing force~\cite{Zhao2015Phys.Rev.C034319,Meng2006Prog.Part.Nucl.Phys470}.
It is found that pairing correlations improve the description of the experimental spectrum and transition probabilities for the yrast band of $^{135}$Nd by considering additional admixtures in the single-particle orbits and altering the orientation of the rotational axis.

The main focus of the present work is to implement the separable form of the Gogny pairing force~\cite{Tian2009Phys.Lett.B44,Niksic2010Phys.Rev.C054318} in the TAC-CDFT.
In comparison with the zero-range monopole pairing force adopted in Ref.~\cite{Zhao2015Phys.Rev.C034319}, the separable pairing force is finite-range and, thus, the problem of an ultraviolet divergence that requires an introduction of a cut-off at large momenta or energies can be avoided.
Meanwhile, due to its separable form, this force requires less cost of the computational time in the practical calculations, as compared to other finite-range pairing forces, such as Gogny force~\cite{Decharge1980Phys.Rev.C1568--1593}.
The developed method of TAC-CDFT with separable force is applied to investigate the yrast band of $^{109}$Ag.
The energy spectra, the relation between spin and rotational frequency, reduced $M1$ and $E2$ transition probabilities are calculated and compared with the available data~\cite{Datta2008Phys.Rev.C21306}.
In particular, the evolution of the rotational axis is discussed in detail.

The paper is organized as follows: after establishing the formalism of the TAC-CDFT with the separable force in Sec.~\ref{sec1}, I discuss in Sec.~\ref{sec2} the numerical details of the method. In Sec.~\ref{sec3}, I compare the calculated results of the yrast band of $^{109}$Ag with corresponding data, and discuss the evolution of the rotational axis. Finally, a summary is given in Sec.~\ref{sec4}.


\section{Theoretical framework}\label{sec1}
The starting point of the point-coupling density functional theory is an effective Lagrangian density of the form
\begin{equation}
  \mathcal{L} = \mathcal{L}^{\mathrm{free}}+\mathcal{L}^{\mathrm{4f}}+\mathcal{L}^{\mathrm{hot}}+\mathcal{L}^{\mathrm{der}}+\mathcal{L}^{\mathrm{em}},
\end{equation}
including the Lagrangian density for free nucleons $\mathcal{L}^{\mathrm{free}}$, the four-fermion point-coupling terms $\mathcal{L}^{\mathrm{4f}}$, the higher order terms $\mathcal{L}^{\mathrm{hot}}$ accounting for the medium effects, the derivative terms $\mathcal{L}^{\mathrm{der}}$ to simulate the finite-range effects that are crucial for a quantitative description of nuclear density distributions, and the electromagnetic interaction terms $\mathcal{L}^{\mathrm{em}}$. The detailed formalism of the point-coupling density functional can be seen, e.g., in Refs.~\cite{Zhao2010Phys.Rev.C054319,Buervenich2002Phys.Rev.C044308,Niksic2008Phys.Rev.C034318}, and the formalism can be easily extended to meson exchange version of the covariant density functional theory~\cite{Long2004Phys.Rev.C034319,Meng2006Prog.Part.Nucl.Phys470}.

To describe the tilted axis rotation of nuclei, as in Ref.~\cite{Zhao2011Phys.Lett.B181}, the Lagrangian is transformed into a frame rotating with a constant frequency in the $xz$ plane,
\begin{equation}
  \bm{\omega} = (\omega_x,0,\omega_z) = (\omega\cos\theta_\omega,0,\omega\sin\theta_\omega),
\end{equation}
where $\theta_\omega$ is the tilted angle between the cranking axis and the $x$-axis.
From the rotating Lagrangian, the equation of motion for nucleons can be derived, which has the form of a Dirac equation
\begin{equation}\label{eq1}
  [\bm{\alpha}\cdotp(\bm{p}-\bm{V})+\beta(m+S)+V-\bm{\omega}\cdotp\hat{\bm{J}}]\psi_k = \varepsilon_k\psi_k,
\end{equation}
where $\hat{\bm{J}}=\hat{\bm{L}}+\frac{1}{2}\hat{\bm{\Sigma}}$ is the total angular momentum, and $S(\bm{r})$ and $V^\mu(\bm{r})$ are the relativistic scalar and vector mean fields, respectively.

For the consideration of pairing correlations, as in Ref.~\cite{Zhao2015Phys.Rev.C034319}, one needs to solve the tilted axis cranking relativistic Hartree-Bogoliubov (RHB) equation for quasiparticles instead of the Dirac equation for nucleons [Eq.~(\ref{eq1})]
\begin{equation}\label{eq2}
  \left(
  \begin{array}{cc}
    h-\bm{\omega}\cdot\hat{\bm{J}}&\Delta\\
    -\Delta^\ast&-h^\ast+\bm{\omega}\cdot\hat{\bm{J}}^\ast
  \end{array}
  \right)
  \left(
  \begin{array}{c}
    U_k\\
    V_k
  \end{array}
  \right) = E_k\left(
  \begin{array}{c}
    U_k\\
    V_k
  \end{array}\right),
\end{equation}
where $h$ is the single-nucleon Dirac Hamiltonian
\begin{equation}
  h_D = \bm{\alpha}\cdot(\bm{p}-\bm{V})+\beta(m+S)+V
\end{equation}
minus the chemical potential $\lambda$, and $\Delta$ is the pairing field.
Here, the pairing field $\Delta$ and the mean fields $S$ and $V^\mu$ in Eq.~(\ref{eq2}) are treated in a unified and self-consistent way.
The scalar and vector fields $S(\bm{r})$ and $V^\mu(\bm{r})$ are determined by
\begin{equation}
  \begin{split}
    S(\bm{r}) &= \alpha_S\rho_S+\beta_S\rho_S^2+\gamma_S\rho_S^3+\delta_S\Delta\rho_S, \\
    V^\mu(\bm{r}) &= \alpha_Vj^\mu_V+\gamma_V(j^\mu_V)^3+\delta_V\Delta j_V^\mu+\tau_3\alpha_{TV}j^\mu_{TV}\\
    &+\tau_3\delta_{TV}\Delta j^\mu_{TV}+eA^\mu,
  \end{split}
\end{equation}
with the densities and currents
\begin{align}
	\rho_S &= \sum_{k>0}{V}_k^\dag\gamma^0 V_k, \\
	j^\mu_V &= \sum_{k>0}\bar{V}_k\gamma^\mu V_k, \\
	j^\mu_{TV} &= \sum_{k>0}\bar{V}_k\gamma^\mu\vec{\tau}V_k,
\end{align}
and the electromagnetic field $eA^\mu$. Here, the sum over $k>0$ corresponds to the well known ``no-sea approximation''~\cite{Niksic2014Comput.Phys.Commun1808}, and $e$ is the electric charge unit vanishing for neutrons.

The matrix element of the pairing field $\Delta$ is
\begin{equation}\label{eq3}
  \Delta_{ab} = \frac{1}{2}\sum_{c,d}\langle ab|V^{pp}|cd\rangle_a\kappa_{cd},
\end{equation}
where $V^{pp}$ is the pairing force, and $\kappa$ is the pairing tensor $\kappa = V^\ast U^T$ determined by the quasiparticle (qp) wavefunctions. In the present work, the separable pairing force is adopted, which reads in the coordinate space,
\begin{equation}\label{Vpp}
  V^{pp}(\bm{r}_1,\bm{r}_2,\bm{r}_1',\bm{r}_2') = G\delta(\bm{R}-\bm{R}')P(\bm{r})P(\bm{r}')\frac{1}{2}(1-P^\sigma).
\end{equation}
Here, $\bm{R} = \frac{1}{2}(\bm{r}_1+\bm{r}_2)$ and $\bm{r}=\bm{r}_1-\bm{r}_2$ denote the center of mass and the relative coordinates respectively, and $P(\bm{r})$ has a Gaussian expression
\begin{equation}
  P(\bm{r}) = \frac{1}{(4\pi a^2)^{3/2}}e^{-r^2/4a^2}.
\end{equation}
The projector $\frac{1}{2}(1-P^\sigma)$ allows only the states with the total spin $S=0$.
The two parameters $G$ and $a$ have been determined in Ref.~\cite{Tian2009Phys.Lett.B44} by fitting to the density dependence of  pairing gaps at the Fermi surface for nuclear matter obtained with the Gogny forces.

By solving Eq.~\eqref{eq2} iteratively, one can obtain the total energy:
\begin{equation}
  E_{\mathrm{tot}} = E_{\mathrm{kin}}+E_{\mathrm{int}}+E_{\mathrm{cou}}+E_{\mathrm{pair}}+E_{\mathrm{c.m.}},
\end{equation}
which includes a kinetic part,
\begin{equation}
  E_{\mathrm{kin}} = \int d^3\bm{r}\sum_{k>0}V^\dag_k[\bm{\alpha}\cdotp\bm{p}+\beta m]V_k,
\end{equation}
an interaction part,
\begin{equation}
  \begin{split}
    E_{\mathrm{int}}& = \int d^3\bm{r}\left\{\frac{1}{2}\alpha_S\rho_S^2+\frac{1}{3}\beta_S\rho_S^3+\frac{1}{4}\gamma_S\rho_S^4+\frac{1}{2}\delta_S\rho_S\Delta\rho_S\right.\\
    & \left.+\frac{1}{2}\alpha_Vj_V^\mu(j_V)_\mu+\frac{1}{2}\alpha_{TV}j^\mu_{TV}(j_{TV})_\mu\right.\\
    & \left.+\frac{1}{4}\gamma_V(j^\mu_V(j_V)_\mu)^2+\frac{1}{2}\delta_V\Delta j^\mu_V(j_V)_\mu\right.\\
    & \left.+\frac{1}{2}\delta_{TV}j_{TV}^\mu\Delta (j_{TV})_\mu\right\},
  \end{split}
\end{equation}
an electromagnetic part,
\begin{equation}
  E_{\mathrm{cou}} = \int d^3\bm{r}\frac{1}{2}eA_0j_p^0,
\end{equation}
a pairing energy part,
\begin{equation}
  E_{\mathrm{pair}} = \frac{1}{2}\mathrm{Tr}[\Delta\kappa],
\end{equation}
and the center-of-mass (c.m.) correction energy $E_{\mathrm{c.m.}}$ accounting for the treatment of center-of-mass motion,
\begin{equation}
  E_{\mathrm{c.m.}} = -\frac{1}{2mA}\langle\hat{\bm{P}}^2_{\mathrm{c.m.}}\rangle,
\end{equation}
where $A$ is the mass number and $\hat{\bm{P}}_{\mathrm{c.m.}} = \sum_i^A\hat{\bm{p}}_i$ is the total momentum in the center-of-mass frame.

The angular momentum components $\bm{J} = (J_x, J_y, J_z)$ in the intrinsic frame at a certain rotational frequency $\hbar\bm{\omega}$ are given by
\begin{align}\label{eq19}
  &J_x = \langle\hat{J}_x\rangle = \sum_{k>0}j_x^{(k)},\\
  &J_y  = 0,\\
  &J_z = \langle \hat{J}_z\rangle = \sum_{k>0}j_z^{(k)},
\end{align}
and the  magnitude of the angular velocity $\bm{\omega}$ is connected to the angular momentum quantum number $I$ by the semiclassical relation $\langle\hat{\bm{J}}\rangle\cdot\langle\hat{\bm{J}}\rangle=I(I+1)$.

The quadrupole moments $Q_{20}$ and $Q_{22}$ are calculated by
\begin{align}
  &Q_{20} = \sqrt{\frac{5}{16\pi}}\langle 3z^2-r^2\rangle,\\
  &Q_{22} = \sqrt{\frac{15}{32\pi}}\langle x^2-y^2\rangle,
\end{align}

The nuclear magnetic moment in units of the nuclear magneton is given by
\begin{equation}\label{eq4}
  \bm{\mu} = \sum_{k>0}\int d^3r\left[\frac{mc^2}{\hbar c}qV_k^\dag(\bm{r})\bm{r}\times\bm{\alpha}V_k+\kappa V_k^\dag(\bm{r})\beta\bm{\Sigma}V_k\right],
\end{equation}
where the charge $q$ ($q_p = 1$ for protons and $q_n = 0$ for neutrons) is given in units of $e$, and $\kappa$ is the free anomalous gyromagnetic ratio of the nucleon $(\kappa_p = 1.793$ and $\kappa_n = -1.913)$. In a semiclassical approximation, the transition probabilities $B(M1)$ and $B(E2)$ values can be derived as
\begin{align}
  &B(M1) = \frac{3}{8\pi}\mu_{\perp}^2=\frac{3}{8\pi}(\mu_x\sin\theta_J-\mu_z\cos\theta_J),\\
  &B(E2) = \frac{3}{8}\left[Q^p_{20}\cos^2\theta_J+\sqrt{\frac{2}{3}}Q_{22}^p(1+\sin^2\theta_J)\right]^2,
\end{align}
where $Q_{20}^p$ and $Q_{22}^p$ correspond to the quadrupole moments of protons.

\section{Numerical details}\label{sec2}
In the present work, the observed yrast band of the odd-$A$ nucleus $^{109}$Ag~\cite{Datta2008Phys.Rev.C21306} is investigated.
The ground state of $^{109}$Ag is associated with the one quasi-proton configuration $\pi g_{9/2}^{-1}$.
However, above $I=21/2\hbar$, two $h_{11/2}$ neutrons are aligned, and this leads to the 3-qp configuration $\pi g_{9/2}^{-1}\otimes\nu h_{11/2}^2$.

In the following, I apply the developed TAC-CDFT with separable pairing force to both the 1-qp and 3-qp configurations.
The point-coupling density functional PC-PK1~\cite{Zhao2010Phys.Rev.C054319} is adopted in the particle-hole channel and the separable pairing force with $G=-738$ MeV fm$^3$ and $a=0.636$ fm~\cite{Tian2009Phys.Lett.B44} are used in the particle-particle channel. The RHB equation~(\ref{eq2}) is solved in a three-dimensional harmonic oscillator basis in Cartesian coordinates with 12 major shells.
By increasing the number of major shell from 12 to 14, the changes of total energy and total angular momentum at the rotational frequency 0.35 MeV are within 0.021$\%$ and 0.020$\%$ respectively.
In the calculations, I follow the method proposed in Ref.~\cite{Zhao2015Phys.Rev.C034319} to trace and block the right qp orbitals to keep the multi-qp configurations unchanged while solving Eq.~\eqref{eq2} iteratively with different $\lambda$ and $\bm{\omega}$ values.

It should be noted that in the present TAC-CDFT with separable pairing force, the time reversal symmetry as well as signature is broken and parity is the only good quantum number.
Therefore, the space of the current Hamiltonian matrix is twice as large as for the corresponding noncranking RHB theory~\cite{Meng2006Prog.Part.Nucl.Phys470,Niksic2014Comput.Phys.Commun1808}.
Furthermore, there are more nonzero matrix elements for pairing field needed to be calculated (see Appendix~\ref{Append1} for details).

\section{Results and discussion}\label{sec3}

\begin{figure}[!htbp]
  \centerline{
  \includegraphics[width=8cm]{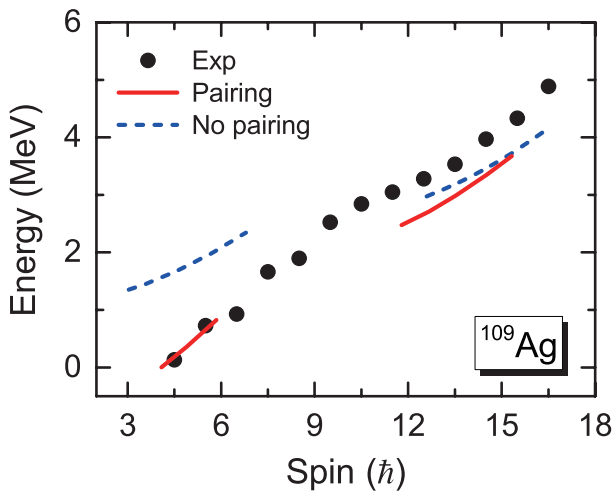}}
  \caption{(Color online)  Rotational excitation energy as a function of the angular momentum in comparison with the data~\cite{Datta2008Phys.Rev.C21306} (solid dots).
  The solid and dashed lines represent the results calculated by TAC-CDFT with and without pairing correlations, respectively.
  Here, the excitation energies are the energy differences with respect to the ground state.}
  \label{fig1}
\end{figure}

In Fig.~\ref{fig1}, the calculated rotational excitation energies for both the 1-qp (lower spin part) and 3-qp (high spin part) configurations are shown in comparison with the data available~\cite{Datta2008Phys.Rev.C21306}.
It is seen that the experimental data are reproduced satisfactorily without any artificial renormalization of the bandhead.
In particular, the description of the energy differences between the bandheads of the 1-qp and 3-qp configurations is improved significantly with the inclusion of pairing correlations.
This feature is very similar to that found in the previous work for the yrast band of $^{135}$Nd~\cite{Zhao2015Phys.Rev.C034319}, where the TAC-CDFT calculations were carried out with an adjusted constant pairing force.
Quantitatively, it seems that the current separable pairing force is not strong enough to reproduce the energy differences between the two bandheads exactly.
However, it is worthwhile to mention that the particle number is not conserved in the present calculations due to the Bogoliubov transformation.
Further consideration of the particle-number restoration could bring more correlations for the 1-qp configuration than the 3-qp one, because the 1-qp configuration has larger pairing gaps.
This would enlarge the energy differences between the two bandheads and, thus, may lead to an improved description of the data.
Further work along this direction is in progress.
An additional reason for the discrepancy in the energy differences of 1-qp and 3-qp configurations could be the missing accuracy of the description of single-particle energies in self-consistent mean field theories.
The transition from the 1-qp and the 3-qp configuration involves a particle-hole excitation. A reduced particle-hole energy in the transition to the two $h_{11/2}$ neutrons could also cause the deviation from the experiment in the energy spectra observed in Fig.~\ref{fig1}.

\begin{figure}[!htbp]
  \centerline{
  \includegraphics[width=8cm]{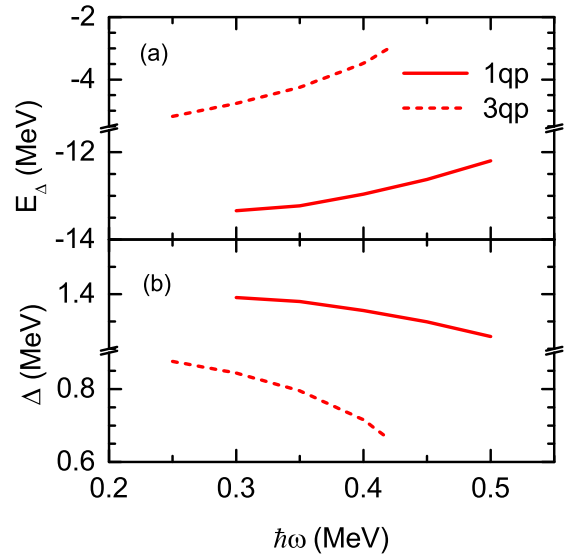}}
  \caption{(Color online) The neutron pairing energy (a) as well as pairing gap (b) as a function of rotational frequency.
  The solid and dashed lines represent the cases of 1-qp and 3-qp configurations respectively.}
  \label{fig2}
\end{figure}

For a better understanding of the results shown in Fig.~\ref{fig1}, the neutron pairing energy and pairing gap for 1-qp and 3-qp bands are shown in Fig.~\ref{fig2}.
Due to the odd proton in $^{109}$Ag blocking the orbital close to the major shell $Z=50$, the pairing correlations for protons vanish.
As one sees from Fig.~\ref{fig2}, the pairing effects of the 3-qp configuration are much weaker than those of the 1-qp configuration.
Moreover, for both configurations, the pairing energy and the pairing gap are decreasing with the rotational frequency, which indicates weaker pairing correlations at high rotational frequency.

\begin{figure}[!htbp]
  \centerline{
  \includegraphics[width=8cm]{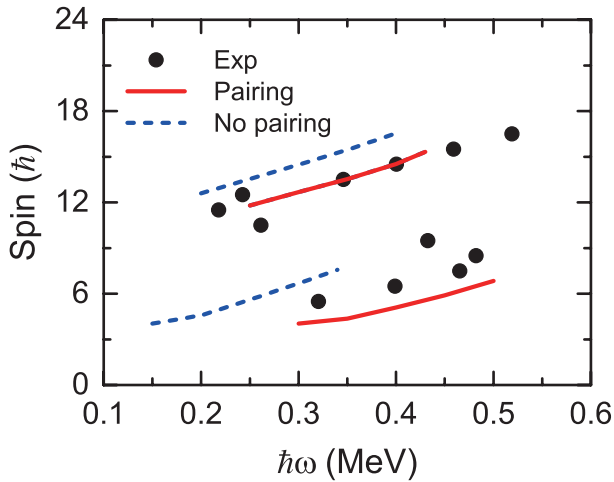}}
  \caption{(Color online) Angular momenta as a function of rotational frequency in comparison with the data~\cite{Datta2008Phys.Rev.C21306} (solid dots). The solid and dashed lines represent the results calculated by TAC-CDFT with and without pairing correlations, respectively.}
  \label{fig3}
\end{figure}

Converged results can not be obtained from $I\sim7.5\hbar$ to $I\sim11.5\hbar$ in Fig.~\ref{fig1} because of the backbending phenomenon observed in this region, where the rotational frequency drops drastically while the angular momentum increases.
This can be clearly seen in Fig.~\ref{fig3}, which depicts the calculated angular momenta as a function of the rotational frequency in compared with the data.
It is well known that such a backbending phenomenon is beyond the scope of a cranking calculation~\cite{Hamamoto1976Nucl.Phys.A15}.
Apart from the backbending region, the angular momenta are reproduced well by the calculations with pairing correlations.
In comparison with the calculated results without pairing correlations, it is clear that the pairing effects tend to slow down the total spin alignments and, thus, bring the results closer to the data for a given rotation frequency.
Moreover, for the 1-qp configuration, converged results can be obtained up to $\hbar\omega\simeq 0.5$ MeV by taking into account the pairing correlations.

\begin{figure}[htbp]
  \centerline{
  \includegraphics[width=8cm]{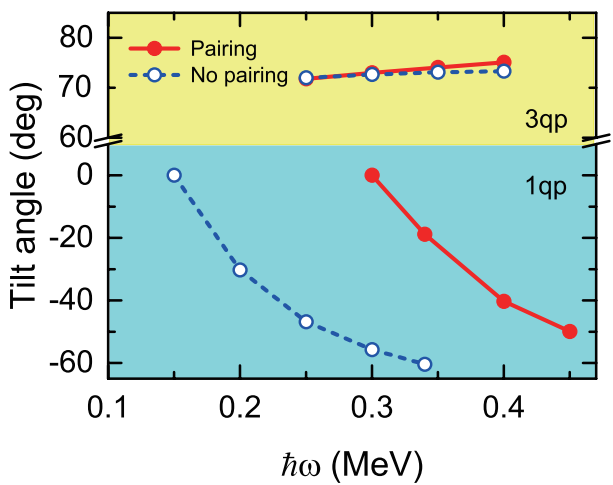}}
  \caption{(Color online) Tilt angles for the 1qp and 3qp configurations as a function of the rotational frequency.
           The full and open circles represent the results calculated by TAC-CDFT with and without pairing correlations, respectively.}
  \label{fig4}
\end{figure}

In order to have a better understanding of the dynamics of the rotational band, it is interesting to study the evolution of the orientation of the rotational axis and the angular momentum vectors of neutrons and protons.
The orientation of the rotational axis can be represented by the so-called tilt angle, which is defined here as the angle between the rotational axis and the long axis, and determined in a self-consistent way by minimizing the total Routhian along the band.
In Fig.~\ref{fig4}, the tilt angles for the 1-qp and 3-qp configurations are shown as a function of the rotational frequency.
The positive and negative values denote a tilt towards the short and the intermediate axes, respectively.
For the 1-qp band, the tilt angles are negative and, thus, the rotational axis is in the long-intermediate ($l$-$i$) plane.
In particular, the orientation of the axis coincides with the $l$ axis at the bandhead because of the quasi-proton in $g_{9/2}$ shell, and is changing gradually towards the $i$ axis when the frequency increases.
For the 3-qp configuration, however, the tilt angles are positive and the axis of rotation lies in the long-short ($l$-$s$) plane.
Moreover, the orientation is close to the $s$ axis due to the appearance of the two aligned quasi-neutrons in the $h_{11/2}$ shell, and does not change much with the increasing frequency.

It is clear in Fig.~\ref{fig4} that pairing correlations influence significantly the orientation of the rotational axis, equivalently the direction of total angular momentum, for the 1-qp band, while having a relatively small impact on the 3-qp band.
This is mainly due to the fact that pairing effects are considerably suppressed by the two aligned quasi-neutrons in the 3-qp band.
To investigate the effect of pairing correlations on the total spin, the magnitudes and directions of the angular momenta for protons and neutrons are studied in detail.

\begin{figure}[htbp]
  \centerline{
  \includegraphics[width=8cm]{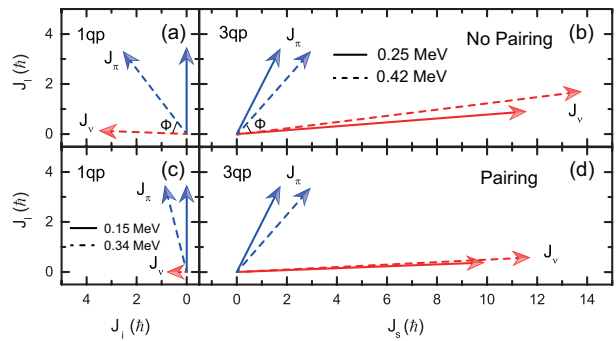}}
  \caption{(Color online) Neutron $\bm{J}_\nu$ and proton $\bm{J}_\pi$ angular momentum vectors for the 1-qp and 3-qp bands with and without pairing correlations.
  The solid and dashed lines represent the results with different values of rotational frequency.
  Note that $\bm{J}_\nu$ for the 1-qp band is negligible at $\hbar\omega=0.15$ MeV.}
  \label{fig5}
\end{figure}

The angular momentum vectors for both the 1-qp and 3-qp bands are depicted in Fig.~\ref{fig5} with and without pairing correlations.
For the 1-qp configuration, the proton angular momentum aligns with the $l$ axis and the neutron one essentially vanishes at the bandhead, i.e., the total spin is dominated by the unpaired quasi-proton in the $g_{9/2}$ shell.
Along the band, the $i$-axis components of the angular momenta for protons and neutrons are increasing due to the coherent collective motion of the nucleons in low-$j$ orbits.
Thus, the proton angular momentum tilts towards the $i$ axis and the neutron one has a non-vanishing contribution.
As a result, the total spin composed by the angular momenta of protons and neutrons also changes away from $l$ axis to $i$ axis due to the collective rotation.

However, this effect is mitigated after the pairing correlations are included, because the pairing correlations provide a remarkable suppression of the angular momentum alignment due to the collective rotation.
This also explains, as depicted in Fig.~\ref{fig4}, the late onset of the decline of the tilt angle with rotational frequencies by the inclusion of pairing correlations.
Note that, in both cases with and without pairing correlations, the $l$-axis projections of the proton angular momenta are nearly constant along the band, since they are determined mainly by the unpaired $\pi g_{9/2}$ quasi-proton.

For the 3-qp band, at the bandhead, the proton angular momenta are mainly from the unpaired $\pi g_{9/2}$ quasi-proton, which align roughly along the $l$ axis, while the neutron ones mainly align with the $s$ axis because of the two $\nu h_{11/2}$ quasi-neutrons.
As the rotational frequency increases, the neutron and proton angular momenta align towards each other, and generate the total angular momenta whose directions are nearly unchanged.
The inclusion of pairing correlations reduces the magnitudes of the neutron angular momenta, while has little influence on the angular momenta of protons.
Consequently, the pairing effects are observed in both the magnitudes (see Fig.~\ref{fig3}) and the directions of the total angular momentum (see Fig.~\ref{fig4}).

\begin{figure}[htbp]
  \centerline{
  \includegraphics[width=8cm]{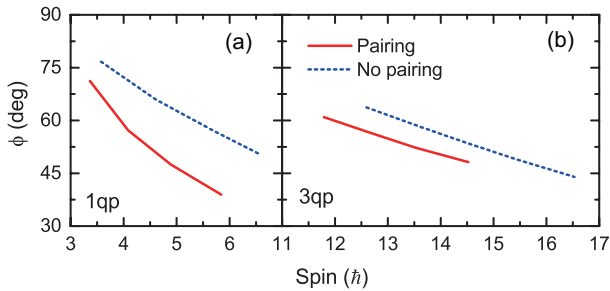}}
  \caption{(Color online) The angle $\phi$ between the proton and neutron angular momenta as a function of the total angular momentum for 1-qp and 3-qp configurations calculated by TAC-CDFT with and without pairing correlations.}
  \label{fig6}
\end{figure}

One can define the angle between the proton and neutron angular momenta, i.e., the angle $\phi$ in Fig.~\ref{fig5}, for a clearer picture of the impact of pairing correlations.
Fig.~\ref{fig6} compares its evolution with respect to the total angular momentum for the cases with and without pairing correlations.
It is clearly seen that this angle $\phi$ is reduced, at each spin, by the inclusion of pairing correlations for both the 1-qp and 3-qp configurations.
Similar phenomenon has also been found in the previous TAC-CDFT calculations for $^{135}$Nd with a monopole pairing force~\cite{Zhao2015Phys.Rev.C034319}.

\begin{figure}[htbp]{}
  \centerline{
  \includegraphics[width=8cm]{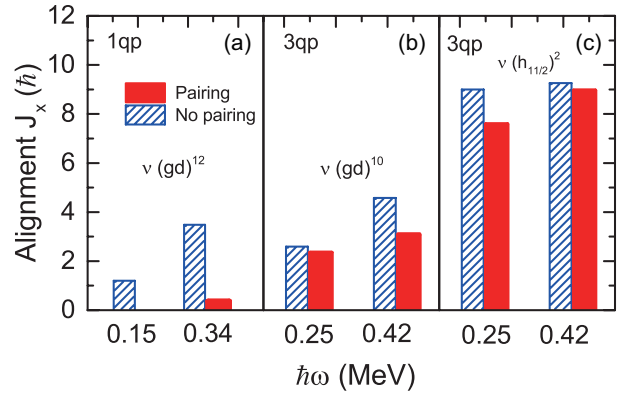}}
  \caption{(Color online) Alignments of the neutron angular momenta on the $x$ axis for both the 1-qp and 3-qp configurations, calculated by TAC-CDFT with and without pairing respectively.
  The numbers below the abscissa denote the rotational frequency at which the plotted alignments have been obtained.}
  \label{fig7}
\end{figure}

To trace the microscopic reason for the pairing effects, it is quite helpful to transform from the qp basis to the canonical basis using  Bogoliubov transformation~\cite{Ring2004}.
In a microscopic picture, the angular momentum comes from all the individual particles.
Here, in Fig.~\ref{fig7}, the neutron angular momentum alignments $J_x$ along the $x$ axis [see Eq.~\eqref{eq19}], i.e., the $i$ axis for the 1-qp band and the $s$ axis for the 3-qp one, are presented as an example. I don't show the proton angular momenta here because the pairing correlations for the protons are actually very weak for $^{109}$Ag (see below).

For the 62 neutrons in the nucleus $^{109}$Ag, the angular momentum is mainly contributed by the 12 neutron particles above the closed $N = 50$ shell.
For the 1-qp configuration, all 12 neutrons are in the $(g_{7/2}d_{5/2})$ shell with low-$j$ values.
Pairing correlations can provide a strong influence on these orbitals and, as a result, the angular momentum alignments along the $x$ axis are significantly reduced.
For the 3-qp configuration, however, there are two neutrons sitting at the high-$j$ orbitals in the $h_{11/2}$ shell, whose angular momentum alignments are hardly influenced by the pairing correlations.
The pairing effects are mainly exhibited by the reduction of the angular momentum alignment for 10 neutrons in the $(g_{7/2}d_{5/2})$ shell.

\begin{figure}[!htbp]
  \centerline{
  \includegraphics[width=8cm]{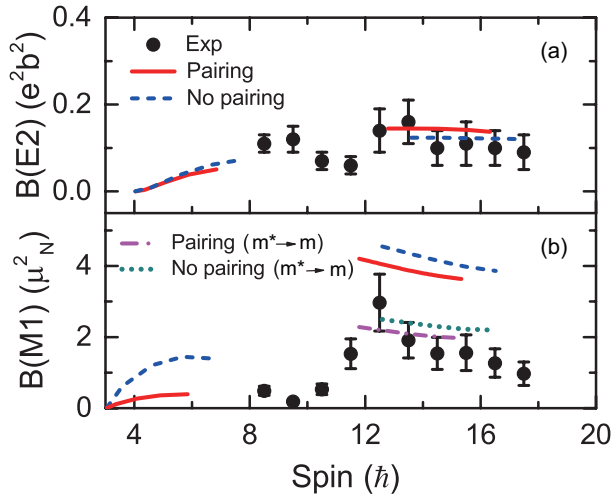}}
  \caption{(Color online) The calculated $B(E2)$ and $B(M1)$ values as a function of the angular momentum in comparison with the data~\cite{Datta2008Phys.Rev.C21306} (solid dots). The solid and dashed lines represent the results calculated by TAC-CDFT with and without pairing correlations, respectively. The $B(M1)$ values are also calculated with scaled Dirac effective mass, and denoted by the dash dotted (with pairing) and dotted lines (without pairing).}
  \label{fig8}
\end{figure}

The electromagnetic transition properties associated with the rotational band are investigated as well.
The theoretical transition probabilities are given in Fig.~\ref{fig8} in comparison with the available data~\cite{Datta2008Phys.Rev.C21306}.
A good agreement with the experiment is achieved by performing the TAC-CDFT calculations with pairing correlations.
The pairing effects are marginal for the $B(E2)$ values, while they are more significant on the $B(M1)$ transitions, whose strengths are reduced by the inclusion of pairing correlations and, thus, approach to the experimental values.
These findings are very similar to those reported in Ref.~\cite{Zhao2015Phys.Rev.C034319} for the yrast band of $^{135}$Nd.
However, the reduction of the $B(M1)$ values caused by the pairing effects in the current work is less remarkable as that in
the previous work of $^{135}$Nd~\cite{Zhao2015Phys.Rev.C034319}, and this is probably related to the fact that the pairing gap obtained here in $^{109}$Ag is smaller.
To clarify this, the average pairing gaps~\cite{Agbemava2014Phys.RevC.054320}
\begin{equation}
	\bar \Delta = \frac{\sum_ku_k\upsilon_k\Delta_k}{\sum_ku_k\upsilon_k}
\end{equation}
are calculated for both protons and neutrons.
Here, $\upsilon_k^2$ denotes the occupation probability of the single-particle state $\psi_k$ in the canonical basis, and it fulfills $\upsilon_k^2+u_k^2=1$, and $\Delta_k$ is the corresponding diagonal matrix element of the pairing field in this basis.

Taking the 3-qp configuration as an example, the proton pairing gap of $^{109}$Ag is almost zero because it is close to
the major shell at $Z=50$ and the blocking of the odd proton reduces pairing correlations in addition.
Similarly, the proton pairing gap of the 3-qp configuration of $^{135}$Nd in Ref.~\cite{Zhao2015Phys.Rev.C034319} is also negligible due to the two aligned quasi-protons in the $h_{11/2}$ shell.
For the neutron pairing gap, however, the one obtained for $^{109}$Ag ($0.87$ MeV) is 20\% smaller than that of $^{135}$Nd ($1.26$ MeV) at the same frequencies.
This might be due to the fact that the two aligned $\nu h_{11/2}$ quasi-neutrons in $^{109}$Ag could suppress the corresponding pairing gap significantly.

The $B(M1)$ values depend strongly on the magnetic moment which can be calculated through Eq.~\eqref{eq4}.
In fact, one can rewrite the first part of Eq.~\eqref{eq4}, i.e., the Dirac magnetic moment $\mu_D$ as
\begin{equation}\label{eq5}
  \mu_D = \sum_{k>0}\int \frac{qm}{m^\ast}\bar{V}_k(\bm{r})[\bm{L}+\bm{\Sigma}]V_k(\bm{r})d\bm{r}.
\end{equation}
Here, $m^\ast$ denotes the Dirac effective mass, and $\bm{L}$ and $\bm{\Sigma}$ are respectively the operators of the orbital and spin angular momentum.
It's known that in the relativistic mean field theory, the effective mass appears too small ($m^\ast\approx 0.58m$), which might be associated with the lack of a fine tuned tensor-coupling vertex.
This leads to a significant enhancement of the Dirac magnetic moment and, thus, additional effects, such as the back-flow effects, are required to describe the magnetic moment properly.
Therefore, here one scales the Dirac effective mass approximately to the nucleon mass by introduction of a factor 0.58~\cite{Zhao2017Phys.Lett.B1}, and it approximately accounts for the back-flow effects, which have been calculated in infinite nuclear matter by a Ward identity~\cite{Bentz1985Nucl.Phys.A593,Arima2011Sci.ChinaPhys.Mech.Astron188,Wei2012Prog.Theor.Phys.Supp.400}.
In such a way, it is found that the calculated $B(M1)$ values are highly consistent with the experimental data.
Moreover, it is noted that this method works not only for the present case but also for the case of the chiral rotation in $^{106}$Rh as reported in Ref.~\cite{Zhao2017Phys.Lett.B1}.

\section{Summary}\label{sec4}

In summary, a separable form of Gogny's pairing force has been implemented in the tilted axis cranking covariant density functional theory for the treatment of pairing correlations.
Contrary to the method of monopole pairing force, the strength of this force is not adjusted to odd-even mass differences.
It is derived from the Gogny force, and in that sense it is universal in the entire periodic table.
This method has been applied to investigate the yrast rotational band of $^{109}$Ag.
The description of the energy spectra, especially the bandhead energies, is improved remarkably when the pairing correlations are considered with the separable pairing force.
For the yrast band of $^{109}$Ag, pairing effects pull down the calculated angular momenta at a certain cranking frequency, and a good agreement with experimental data is thus achieved.
Along the band, the rotational axis undergoes a sharp transition from the long-intermediate plane to the long-short one, which is in accompany with the change of the configurations from 1-qp to 3-qp.
The electromagnetic transition strengths $B(E2)$ and $B(M1)$ are also well reproduced.
In particular, for the 3-qp band, the $B(M1)$ values are reduced by the inclusion of pairing correlations, and this should be  connected with the reduction of the angle between the proton and neutron angular momenta by pairing correlations.

\begin{acknowledgments}
  The author thanks to J. Meng, P. Ring, L. S. Song, S. Q. Zhang and P. W. Zhao for helpful discussions and careful readings of the manuscript.
  This work is supported in part by the Major State 973 Program of China (Grant No. 2013CB834400), the National Natural Science Foundation of China (Grants No. 11335002, No. 11375015, No. 11461141002, No. 11621131001).
\end{acknowledgments}

\begin{appendix}
\section{Calculation of pairing matrix elements}\label{Append1}
The harmonic oscillator basis one used to solve the tilted axis cranking RHB equation [Eq.~(\ref{eq2})] reads
\begin{equation}\label{A3}
  |n_xn_yn_z;i = +\rangle = |n_xn_yn_z\rangle\frac{i^{n_y}}{\sqrt{2}}[|\uparrow\rangle-(-1)^{n_x}|\downarrow\rangle],
\end{equation}
\begin{equation}\label{A4}
  \begin{split}
    |n_xn_yn_z;i = -\rangle &= |n_xn_yn_z\rangle(-1)^{n_x+n_y+1}\\
    &\times\frac{i^{n_y}}{\sqrt{2}}[|\uparrow\rangle+(-1)^{n_x}|\downarrow\rangle].
  \end{split}
\end{equation}
Here, $|n_xn_yn_z\rangle$ is the harmonic oscillator wave function in Cartesian coordinates, and $n_x, n_y, n_z$ are the corresponding quantum numbers. The labels $i=+$ and $i=-$ represent the states with positive and negative $x$-simplex, respectively, and for simplicity, they are respectively abbreviated below as, $|\alpha\rangle$ and $|\bar{\beta}\rangle$.

Based on this harmonic oscillator basis, the antisymmetric pairing matrix elements $\langle ab|V^{pp}|cd\rangle_a$ in Eq.~(\ref{eq3}) can be calculated,
where the pairing force $V^{pp}$ [Eq.~(\ref{Vpp})] can be written as
\begin{equation}\label{A2}
  \begin{split}
    V^{pp}(\bm{r}_1,\bm{r}_2;\bm{r}_1',\bm{r}_2') & = G\delta(\bm{R}-\bm{R}')P(\bm{r})P(\bm{r}')\frac{1}{2}(1-P^\sigma)\\
    &\equiv W(\bm{r}_1,\bm{r}_2;\bm{r}_1',\bm{r}_2')\frac{1}{2}(1-P^\sigma).
  \end{split}
\end{equation}
These are four types of such matrix elements non-vanishing,
\begin{align}
  &\langle\alpha\bar{\beta}|V^{pp}|\gamma\bar{\delta}\rangle_a = \langle\alpha\bar{\beta}|W\frac{1}{2}(1-P^\sigma)|\gamma\bar{\delta}\rangle_a,\label{A5}\\
  &\langle\alpha\beta|V^{pp}|\gamma\delta\rangle_a = \langle\alpha\beta|W\frac{1}{2}(1-P^\sigma)|\gamma\delta\rangle_a,\label{A6}\\
  &\langle\alpha\beta|V^{pp}|\bar{\gamma}\bar{\delta}\rangle_a = \langle\alpha\beta|W\frac{1}{2}(1-P^\sigma)|\bar{\gamma}\bar{\delta}\rangle_a,\label{A7}\\
  &\langle\bar{\alpha}\bar{\beta}|V^{pp}|\bar{\gamma}\bar{\delta}\rangle_a = \langle\bar{\alpha}\bar{\beta}|W\frac{1}{2}(1-P^{\sigma})|\bar{\gamma}\bar{\delta}\rangle_a\label{A8},
\end{align}
because the operator $\frac{1}{2}(1-P^\sigma)$ projects onto the $S=0$ spin-singlet product state
\begin{equation}
  \begin{split}
    |\gamma\bar{\delta}\rangle_{S=0} =& -|\bar{\delta}\gamma\rangle_{S=0} =\frac{1}{2}i^{n_y^\gamma+n_y^\delta}(-1)^{n_y^\delta+1}\\
    &\times\delta_{n_x^\gamma+n_x^\delta,\mathrm{even}}[|\uparrow\downarrow\rangle-|\downarrow\uparrow\rangle]|n^\gamma n^\delta\rangle,
  \end{split}
\end{equation}
\begin{equation}
  \begin{split}
    |\gamma\delta\rangle_{S=0} = &-|\delta\gamma\rangle_{S=0}=\frac{1}{2}i^{n_y^\gamma+n_y^\delta}(-1)^{n_x^\delta+1}\\
    &\times\delta_{n_x^\gamma+n_x^\delta,\mathrm{odd}}[|\uparrow\downarrow\rangle-|\downarrow\uparrow\rangle]|n^\gamma n^\delta\rangle.
  \end{split}
\end{equation}

It should be noted that in the noncranking or the principal-axis-cranking RHB theory, the terms in Eqs.~\eqref{A6}-\eqref{A8} vanish because of the spatial symmetries fulfilled by the nuclear density distribution. As a result, only the matrix elements in Eq.~\eqref{A5} needs to be calculated.
However, in the present tiled axis cranking case, one has only the space-reflection symmetry for nuclear density distribution.
Therefore, all the four kinds of matrix elements in Eqs.~\eqref{A5}-\eqref{A8} could be nonzero.

Since the final forms of Eqs.~\eqref{A7} and \eqref{A8} are the same as that of Eq.~\eqref{A6}, here I give only the detailed derivations of Eqs.~\eqref{A5} and \eqref{A6},
\begin{equation}
  \begin{split}
    \langle \alpha\bar{\beta}|V^{pp}|\gamma\bar{\delta}\rangle_a = & (-i)^{n_y^\alpha-n_y^\beta}\delta_{n_x^\alpha+n_x^\beta,\mathrm{even}}i^{n_y^\gamma-n_y^\delta}\\
    &\times \delta_{n_x^\gamma+n_x^\delta,\mathrm{even}}\langle n^\alpha n^\beta|W|n^\gamma n^\delta\rangle,
  \end{split}
\end{equation}
\begin{equation}
  \begin{split}
    \langle\alpha\beta|V^{pp}|\gamma\delta\rangle_a = & i^{n_y^\alpha+n_y^\beta}(-1)^{n_x^\beta}\delta_{n_x^\alpha+n_x^\beta,\mathrm{odd}}i^{n_y^\gamma+n_y^\delta}(-1)^{n_x^\delta}\\
    &\times\delta_{n_x^\gamma+n_x^\delta,\mathrm{odd}}\langle n^\alpha n^\beta|W|n^\gamma n^\delta\rangle.
  \end{split}
\end{equation}
For the evaluation of
\begin{equation}
  \begin{split}
    \langle n^\alpha n^\beta|W|n^\gamma n^\delta\rangle = &\int\phi_{n_\alpha}(\bm{r}_1)\phi_{n_\beta}(\bm{r}_2)W(\bm{r}_1,\bm{r}_2;\bm{r}_1',\bm{r}_2')\\
    &\times \phi_{n_\gamma}(\bm{r}_1')\phi_{n_\delta}(\bm{r}_2')d^3r_1d^3r_2d^3r_1'd^3r_2',
  \end{split}
\end{equation}
it can be decomposed into the $x$, $y$ and $z$ parts
\begin{equation}
  \langle n^\alpha n^\beta|W|n^\gamma n^\delta\rangle=GW_xW_yW_z.
\end{equation}
For example, the $x$ component reads
\begin{equation}
  \begin{split}
    W_x = &\int\phi_{n_x^\alpha}(x_1,b_x)\phi_{n_x^\beta}(x_2,b_x)P(x)\delta(X-X')P(x')\\
    &\times\phi_{n_x^\gamma}(x_1',b_x)\phi_{n_x^\delta}(x_2',b_x)dx_1dx_2dx_1'dx_2'.
  \end{split}
\end{equation}
Such an integral can be evaluated by means of the Talmi-Moshinsky transformation and the generating function for the Hermite polynomials~\cite{Niksic2014Comput.Phys.Commun1808}.
Finally, the pairing matrix elements can be represented by a sum of a few separable terms in a basis of three-dimensional harmonic oscillator,
\begin{subequations}\label{A9}
  \begin{equation}
    \langle\alpha\bar{\beta}|V^{pp}|\gamma\bar{\delta}\rangle_a = G\sum_{N_x=0}\sum_{N_y=0}\sum_{N_z=0}\left(V_{\alpha\bar{\beta}}^{N_xN_yN_z}\right)^\ast V_{\gamma\bar{\delta}}^{N_xN_yN_z},
  \end{equation}
  \begin{equation}
    \langle\alpha\beta|V^{pp}|\gamma\delta\rangle_a = G\sum_{N_x=0}\sum_{N_y=0}\sum_{N_z=0}\left(V_{\alpha\beta}^{N_xN_yN_z}\right)^\ast V_{\gamma\delta}^{N_xN_yN_z}.
  \end{equation}
\end{subequations}
Here, the single-particle matrix elements read
\begin{align}
  V_{\gamma\bar{\delta}}^{N_xN_yN_z} &=\delta_{n_x^\gamma+n_x^\delta,\mathrm{even}} i^{n_y^\gamma-n_y^\delta}W^{N_x}_{n_x^\gamma n_x^\delta}W^{N_y}_{n_y^\gamma n_y^\delta}W^{N_z}_{n_z^\gamma n_z^\delta},\\
  V^{N_xN_yN_z}_{\gamma\delta} &= \delta_{n_x^\gamma+n_x^\delta,\mathrm{odd}} i^{n_y^\gamma+n_y^\delta}(-1)^{n_x^\delta} W^{N_x}_{n_x^\gamma n_x^\delta}W^{N_y}_{n_y^\gamma n_y^\delta}W^{N_z}_{n_z^\gamma n_z^\delta},
\end{align}
and the factors $W^{N_\mu}_{n_\mu^\gamma n_\mu^\delta}$ are given by
\begin{equation}
  W^{N_\mu}_{n_\mu^\gamma n_\mu^\delta}=\frac{1}{b_\mu} M^{nN_\mu}_{n_\mu^\gamma n_\mu^\delta}I_n(\alpha_\mu),\quad n = N_\mu-n_\mu^\gamma-n_\mu^\delta,
\end{equation}
where $b_\mu$ is the harmonic oscillator length, and $\alpha_\mu = a/b_\mu$.
$M^{nN_\mu}_{n_\mu^\gamma n_\mu^\delta}$ denotes the Talmi-Moshinsky bracket,
\begin{equation}
  \begin{split}
    M_{n_\mu^\gamma n_\mu^\delta}^{nN_\mu} &= \sqrt{\frac{n_\mu^\gamma! n_\mu^\delta!}{n!N_\mu!}}\sqrt{\frac{1}{2^{N_\mu+n}}}\delta_{n_\mu^\gamma+n_\mu^\delta,n+N_\mu}\\
    &\sum_m(-1)^{n+m}\times
    \left(\begin{array}{c}
      N_\mu\\
      N_\mu-n+m
    \end{array}\right)
    \left(\begin{array}{c}
      n\\
      m
    \end{array}\right),
  \end{split}
\end{equation}
and $I(\alpha_\mu)$ reads
\begin{equation}
  \begin{split}
     I_n(\alpha_\mu) = &\delta_{n,\mathrm{even}}\frac{(-1)^{n/2}}{(2\pi)^{1/4}}\frac{\sqrt{n!}}{2^{n/2}(n/2)!}\\
     &\times\left(\frac{1}{1+\alpha_\mu^2}\right)^{1/2}\left(\frac{1-\alpha^2_\mu}{1+\alpha_\mu^2}\right)^{n/2}.
  \end{split}
\end{equation}

Through Eqs.~\eqref{eq3} and \eqref{A9}, one can finally get the matrix elements of the pairing field
\begin{subequations}
  \begin{equation}
    \Delta_{\alpha\bar{\beta}} = G\sum_{N_x=0}\sum_{N_y=0}\sum_{N_z=0}\left(V_{\alpha\bar{\beta}}^{N_xN_yN_z}\right)^\ast P_{N_xN_yN_z},
  \end{equation}
  \begin{equation}
    \Delta_{\alpha\beta} = G\sum_{N_x=0}\sum_{N_y=0}\sum_{N_z=0}\left(V_{\alpha\beta}^{N_xN_yN_z}\right)^\ast P_{N_xN_yN_z},
  \end{equation}
\end{subequations}
with the coefficients
\begin{equation}
  \begin{split}
    P_{N_xN_yN_z} &= \sum_{\gamma\bar{\delta}>0}V_{\gamma\bar{\delta}}^{N_xN_yN_z}\kappa_{\gamma\bar{\delta}},\\
    &\mathrm{or}\\
    P_{N_xN_yN_z} &= \sum_{\gamma\delta>0}V^{N_xN_yN_z}_{\gamma\delta}\kappa_{\gamma\delta}.
  \end{split}
\end{equation}

\end{appendix}
%

\end{document}